\documentclass[aps, prl, amsmath,amssymb, reprint, showpacs,superscriptaddress,groupedaddress, nofootinbib ]{revtex4-1}

\usepackage{graphicx}
\usepackage{dcolumn}
\usepackage{bm}
\usepackage{amsmath}
\usepackage{color}

\usepackage{dcolumn}  
\usepackage{amsmath}
\usepackage{multirow}
\usepackage{graphicx}   
\usepackage{subfigure}
\usepackage{comment}
\usepackage{color}
\usepackage[colorlinks,bookmarks=false,citecolor=blue,linkcolor=red,urlcolor=blue]{hyperref}
\usepackage{nicefrac}
\usepackage{soul}
\usepackage{slashbox}
\usepackage{bm,array}
\usepackage{tikz}
\usepackage{pifont}
\usepackage{gensymb}

\newcolumntype{C}{>{\centering\arraybackslash}p{1em}}

\begin{document}
	
	\title{High Performance Ternary Alkali Nitrides for Renewable Energy Applications }
	\author{Jiban Kangsabanik and Aftab Alam }
	\email{aftab@iitb.ac.in}	
	\affiliation{$^1$Department of Physics, Indian Institute of Technology Bombay, Powai, Mumbai 400076, India}
	
	\begin{abstract}
		Rapid decline in fossil fuel energy necessitates the immediate need for renewable energy resources. Here, we report a previously unexplored class of nitrides AMN$_2$ (A=Na, K, Rb, Cs, M=V, Nb, Ta) keeping renewable energy applications in mind.  Using a detailed structure and stability analysis using first principles simulation, we discovered twelve such compounds (few of which are already synthesized before), which are chemically, mechanically and dynamically stable. These twelve compounds were then evaluated for their suitability for three renewable energy applications, (i) photovoltaics, (ii) water splitting, and (iii) thermoelectrics. Careful analysis of electronic structure reveals high optical transition strength resulting in sharp rise in absorption. This in turn yields high short circuit current and hence excellent solar efficiency for few compounds namely CsVN$_2$ and RbVN$_2$. Along with excellent absorption quality, some compounds show favorable band edge positions compared to water redox levels and hence are promising as photoelectrodes in photo(electro)chemical water splitting devices. Mixture of flat and dispersive bands in the band structure yields both high Seebeck and electrical conductivity, thus excellent power factor for six compounds. Simulated  lattice thermal conductivity shows moderate to ultralow values and thus the possibility of achieving high thermoelectric figure of merit (ZT), even at lower temperatures. From the experimental perspective, we discuss the possible challenges that may arise while utilizing these compounds for the desired applications, and suggest possible pathways to overcome them. We believe such theoretical prediction of promising materials are extremely useful for new materials discovery and anticipate rapid response from the experimental community.

	\end{abstract}

	\maketitle
	
	{\bf Introduction:} Diminishing fossil fuel energy sources bring us in a situation with immediate need of renewable energy resources. Solar energy (Photovoltaics), waste heat recovery (Thermoelectrics), Photochemical water splitting are some of the well known energy sources showing promise towards a sustainable energy usage for the future generation.\cite{chu2017path, montoya2017materials, pham2017modelling, polman2016photovoltaic} Till now, the above mentioned energy resources cannot compete with the well established fossil fuel energy industry mainly due to higher maintenance cost of the former. This problem can be solved in two different ways, either by improving the efficiency and cutting the processing cost of already well-known materials, which seems to be reaching a saturation point. The other way is to find novel materials, which can be better suited for those applications. Ever increasing computational power along with already developed sophisticated theories make the task of discovering novel materials a lot easier, which is much more cumbersome and expensive for experimentalists alone. Density functional theory (DFT)\cite{kohn1965self} is a very successful method which solves the many body Hamiltonian to gain insight of various material properties, requiring only atomic information as input. High-throughput calculations for screening a large database of compounds keeping specific applications in mind, is very popular now-a-days. This not only evaluates the already existing compounds, but also predicts various well suited materials which were previously unknown experimentally.\cite{curtarolo2013high,mounet2018two} 
	
	Here, we investigated a rarely explored class of compounds AMX$_2$ (A=Li, Na, K, Rb, Cs, M=V, Nb, Ta, X=N, P, As, Sb). Here A are alkali metals with $`$+1' charge state, whereas M are group-V transition elements holding  $`$+5' charge states and X are pnictogen anions with charge state $`$-3', thus balancing the charge neutrality. As far as Nitrides are concerned, binary nitrides are well known to show potential both from application and fundamental physics perspective.\cite{al2016two,dev2008defect} Ternary nitrides (AMN$_2$), however,  are relatively less explored\cite{szymanski2018dynamical, pandey2017ii, martinez2017synthesis, hinuma2016discovery, han2018recent, balogun2017updates}. In the above mentioned class of nitrides (AMN$_2$; A=Li, Na, K, Rb, Cs, M=V, Nb, Ta), few were experimentally synthesized in the past\cite{rauch1992ambient, jacobs1989synthese, jacobs1993synthesis, jacobs1993synthesis2}. This encourages us to screen the whole series. We first verify the stability of these compounds via considering different decomposition pathways. Next, we proceed with detailed electronic structure calculation of the stable compounds, and evaluate their potential for various renewable energy applications by simulating a number of carefully designed and well known descriptors. 
	
	\begin{figure}[t]
		\centering
		\includegraphics[width=1\linewidth]{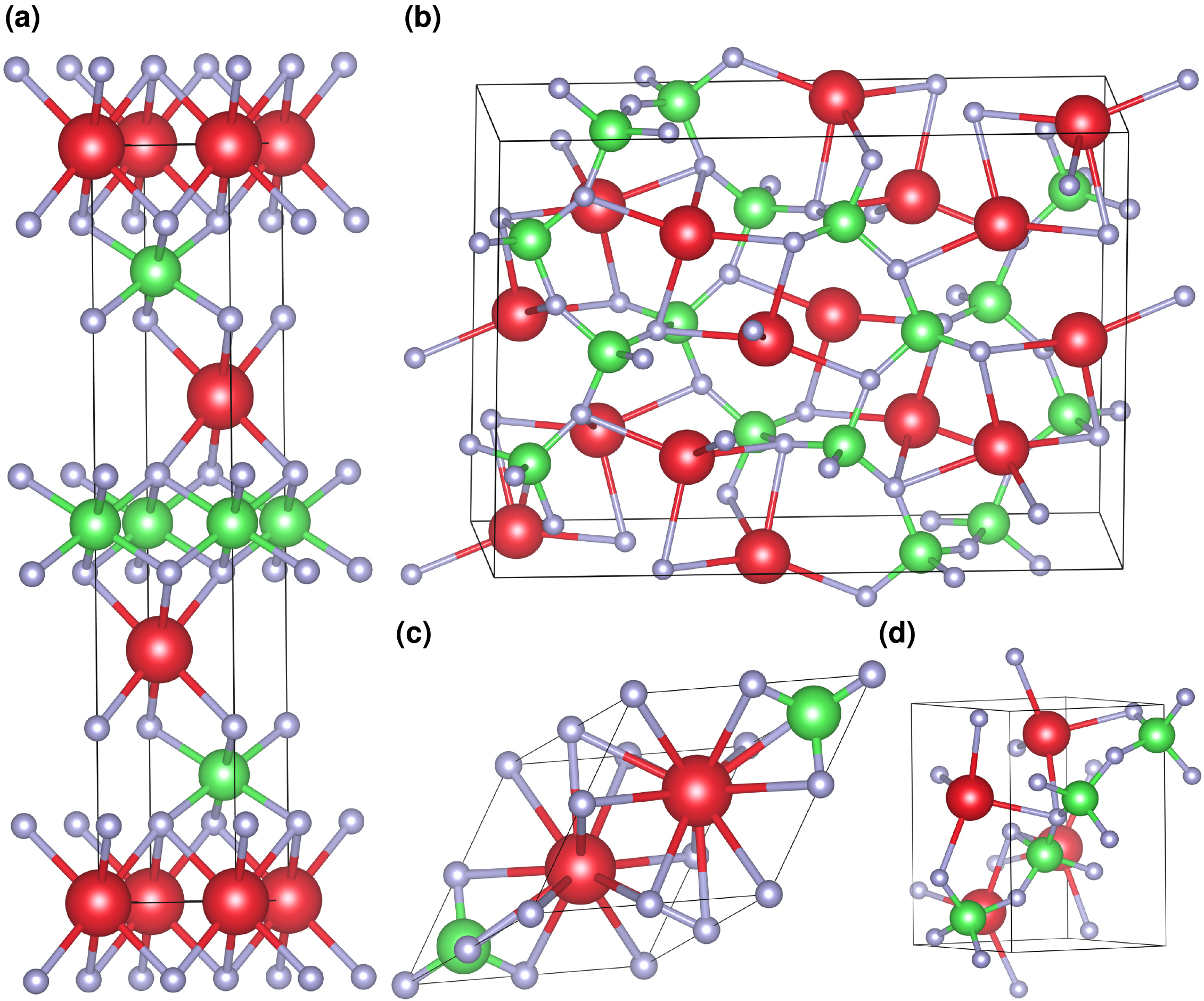}
		\caption{Various crystal structures of stable AMN$_2$ (A=Na, K, Rb, Cs, M=V, Nb, Ta) compounds with space group (a) R-3m:H, (b) Pbca, (c) F-d3m, and (d) Pna2$_1$. Here A, M and N atoms are shown using red, green and grey spheres respectively.  }
		\label{fig:1}
	\end{figure}

\begin{figure}[t]
		\centering
		\includegraphics[width=0.90\linewidth]{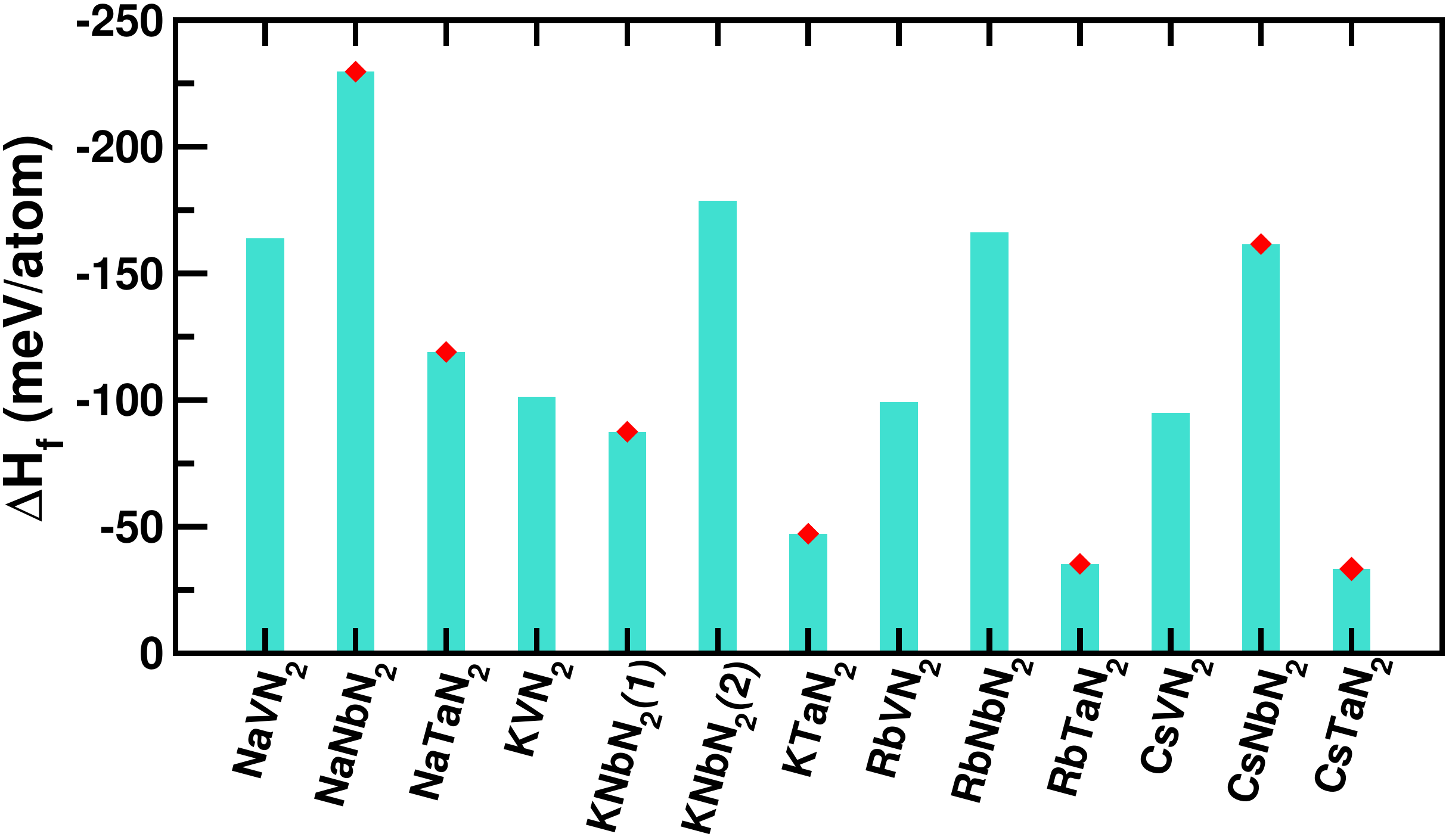}
		\caption{Formation enthalpies for the stable nitride compounds against their respective most stable decomposition pathway. The ones with the red diamond on top are experimentally reported compounds.\cite{rauch1992ambient, jacobs1989synthese, jacobs1993synthesis, jacobs1993synthesis2}  }
		\label{fig:2}
	\end{figure}

	{\bf Stability and Structural details: }
	Depending on the constituent elements, AMX$_2$ class of compounds are, in general, reported to crystallize in different crystal structures.\cite{martinez2017synthesis} Keeping that in mind, for our AMX$_2$ series of compounds, we simulated 10 different crystal structures (space groups) namely Fm-3m, F-43m, F-d3m, I-42d, Pbca, Pnma, Pna2$_1$, P-3m1, R-3m:R and R-3m:H. We have relaxed all the compounds from these starting structures and calculated the decomposition enthalpies against possible elemental and binary phases found in Inorganic Crystal Structure Database (ICSD). We have found that only Nitride compounds are stable against the most favorable decomposition pathway AMN$_2$ $\rightarrow$ AN$_3$ + M$_x$N$_y$ (VN, NbN, Ta$_3$N$_5$)), and have finite band gaps. All the other pnictides (except three phosphide compounds showing metallic behaviour with PBE exchange correlation functional) came out to be unstable (see Sec. S2(A) of  supplementary material (SM)\cite{supplement} for various decomposition pathways including their energetics. In addition, we have also searched for possible higher order compounds A$_p$M$_q$N$_r$ (A=Li, Na, K, Rb, Cs, M=V, Nb, Ta), $p,q,r$ taken from 1-10, available in ICSD. We found 4 such compounds namely Li$_7$VN$_4$, Li$_7$NbN$_4$, Li$_7$TaP$_4$, NaTa$_3$N$_5$. For these compounds, we have also calculated the decomposition enthalpies against possible elemental and binary phases and compared with corresponding AMX$_2$ compounds. We found Li$_7$VN$_4$, Li$_7$TaN$_4$ to be energetically more stable than LiVN$_2$, LiTaN$_2$ respectively whereas, LiNbN$_2$ and NaTaN$_2$ turn out to be relatively more stable than Li$_7$NbN$_4$ and NaTi$_3$N$_5$. For few other nitrides, we noticed close formation enthalpies for more than one crystal structures. For such compounds we chose the ground state structure based on our experience from already experimentally reported compounds (NaTaN$_2$, NaNbN$_2$, KTaN$_2$, KNbN$_2$, RbTaN$_2$, CsTaN$_2$, CsNbN$_2$)\cite{rauch1992ambient, jacobs1989synthese, jacobs1993synthesis, jacobs1993synthesis2} in this class, and the corresponding simulated dynamical stability (phonon dispersion). We have tabulated the chosen crystal structures and decomposition enthalpies against the most favorable decomposition pathway, for the stable compounds, in Table S1 of SM.\cite{supplement} Figure~\ref{fig:2} shows the decomposition enthalpies of all the twelve nitrides. For KNbN$_2$, space group Pbca has somewhat higher (more negative) chemical formation enthalpy than the experimentally reported Fd-3m phase. That is why, we have considered both the structures for this compound, labelled as KNbN$_2$(1) and KNbN$_2$(2) for Fd-3m and Pbca phase respectively. We have also checked mechanical stability of all these compounds by verifying Born-Huang mechanical stability criteria.\cite{born1954dynamical} Except LiNbN$_2$, all the compounds came out to be mechanically stable. Details of the computational procedure for decomposition enthalpy, mechanical stability and further informations about the competing phases can be found in SM.\cite{supplement}

	\begin{figure*}[t]
		\centering
		\includegraphics[width=1\linewidth]{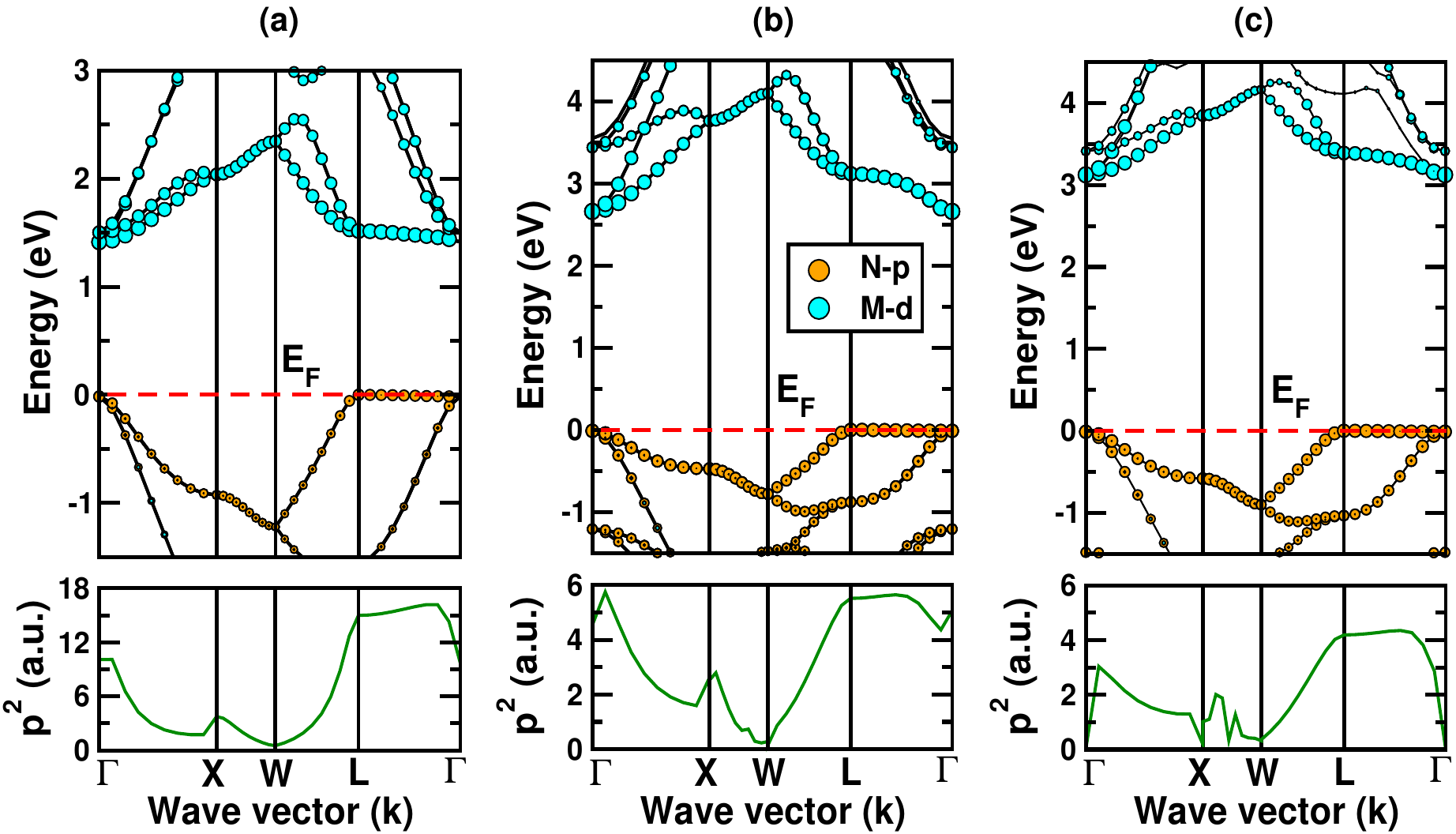}
		\caption{Electronic band structure and square of dipole transition matrix elements (p$^2$) between VBM and CBM (aka optical transition probability) for cubic (F-d3m) (a) CsVN$_2$ (b) CsNbN$_2$ (c) CsTaN$_2$ compounds. Here Nitrogen p and M-d (M=V, Nb, Ta) orbitals are depicted by orange and turquoise coloured circles respectively.}
		\label{fig:3}
	\end{figure*}	
	Our simulation shows that those twelve nitrides stabilize in four different space groups, as depicted in Fig.~\ref{fig:1}. They are R-3m:H (NaNbN$_2$, and NaTaN$_2$),  Pbca (KNbN$_2$, KTaN$_2$, RbVN$_2$, RbNbN$_2$ and RbTaN$_2$), Pna2$_1$ (NaVN$_2$, and KVN$_2$), and F-d3m (KNbN$_2$, CsVN$_2$, CsNbN$_2$ and CsTaN$_2$). In case of R-3m:H, alkali atoms sit at $`$3b' Wyckoff site, whereas transition element (Nb or Ta) sits at $`$3a' Wyckoff site forming an octahedral coordination with Nitrogen (N) atoms occupying $`$6c' site. In the cubic (F-d3m) structure M (V, Nb, Ta) atoms ($`$8a' Wyckoff site) are tetrahedrally coordinated by N ($`$16c'), with alkali atoms ($`$8b') at twelve coordinated position. Space group Pbca (orthorhombic) can be seen as a lower symmetry version of the cubic structure discussed above with all elements occupying different $`$8c' Wyckoff positions. In case of NaVN$_2$, and KVN$_2$, both alkali (Na, K) and V are tetrahedrally coordinated by N with all the elements sitting at distinct $`$4a' sites. Here the transition metals (A) are in a +5 valence state and along with the alkali (A=+1) ion provide 6-electrons to donate to the 2 N (-3 state) in the formula unit, hence making a completely ionic picture. In other words, we have 1 N-2s and 3 N-2p orbital derived bands per N or in total 8 bands or space for 16 electrons per formula unit and these are provided by the A (+1) and VB (+5) element + 2 N (+5*2=10). So, the gap
	should be between N-p and transition metal-d bands, which we discuss later. Our calculation of Born effective charges (see Table S5 in SM\cite{supplement}) show mixed covalent ionic nature of bonding in some compounds with others showing ionic bonding characteristics. By comparing the compounds in their respective structures, we see that ionicity increases with decreasing atomic mass of alkali atoms (from Cs to Na), whereas heavier transition elements show more ionic nature of bonding (Ta $>$ V). Ta compounds mostly show ionic bonding with exception being NaTaN$_2$, forming in a layered structure. Vanadium compounds mainly show mixed covalent ionic characteristics in V-N bonding.  After finding out the most stable AMN$_2$ structures, we then proceed to calculate the electronic structure and other relevant properties for renewable energy applications.

	\begin{figure*}[t]
		\centering
		\includegraphics[width=1\linewidth]{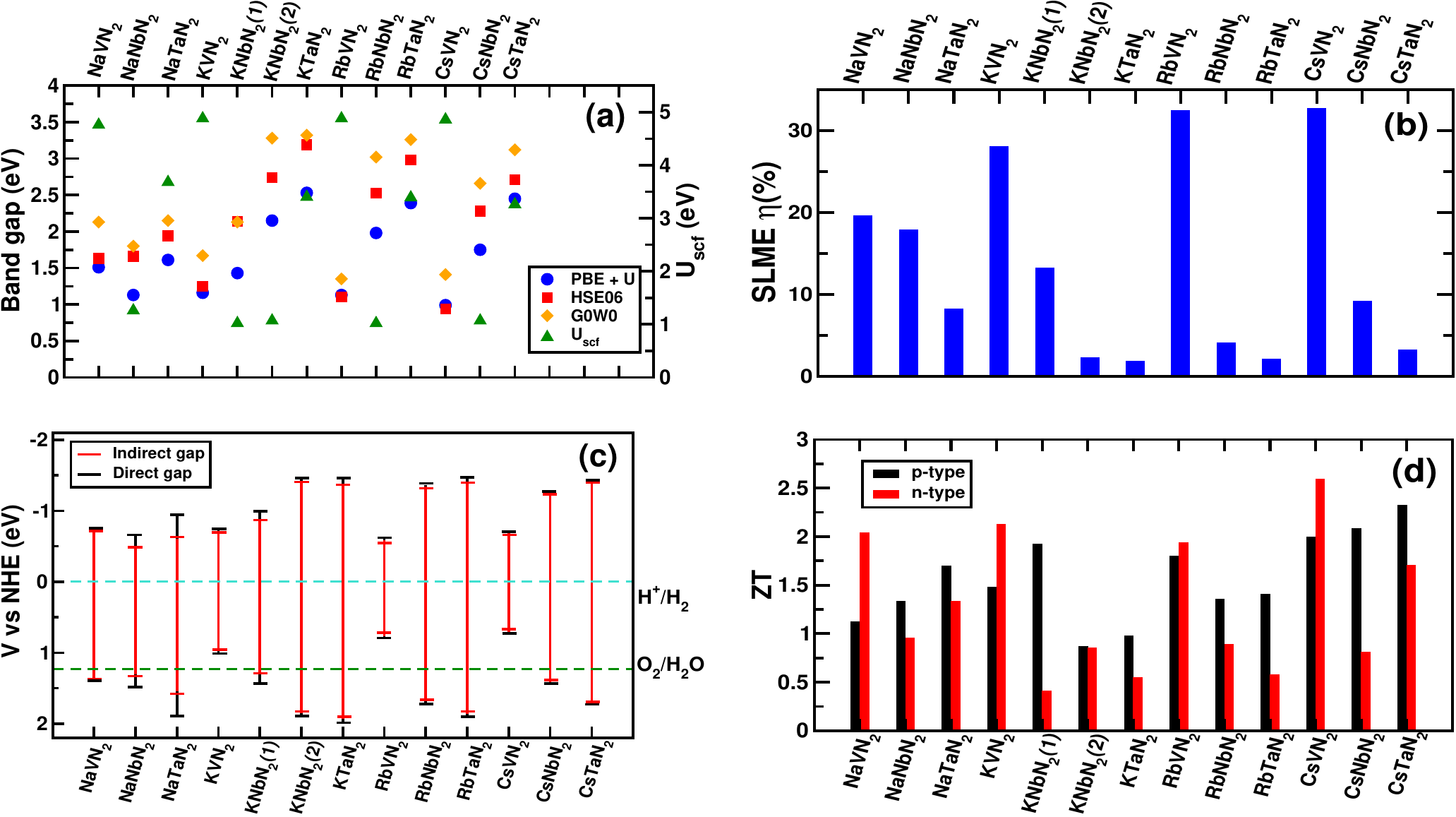}
		\caption{ For stable nitrides, (a) Band gaps using different exchange correlation functionals and self-consistent Hubbard $`$U'$_{scf}$ value, (b) SLME, (c) Band edge positions with respect to water redox levels and (d) maximum achievable ZT value with  n-type and p-type doping (at 600 K).   }
		\label{fig:4}
	\end{figure*}

	{\bf Electronic structure: }
	As these compounds contain transition elements, electron correlation can play an important role in predicting their electronic properties. Here, we have used Perdew-Burke-Ernzerhof (PBE) exchange correlation functional with Hubbard $`$U' correction for transition metal d-orbitals, for primary electronic structure calculations.\cite{perdew1996generalized} We have calculated the $`$U' parameters self-consistently for each compound.\cite{cococcioni2005linear} Details about our calculation can be found in SM.\cite{supplement} In Figure~ \ref{fig:3}, we show electronic band structure along with the optical transition probability for CsMN$_2$ (M=V, Nb, Ta) compounds as representative systems. These compounds show slightly indirect band gap with optically allowed direct gap positioned nearby. For improved accuracy, the band gaps are calculated using quasiparticle G$_0$W$_0$ method, starting from HSE06 ground state.\cite{krukau2006influence} Figure~ \ref{fig:4}(a) shows the band gaps calculated with PBE+U, HSE06, and G$_0$W$_0$, exchange correlation. For PBE+U, the value of U=U$_{scf}$ used for each compound is calculated self-consistently, as also shown by triangle up symbol in Figure~ \ref{fig:4}(a). G$_0$W$_0$ estimated band gap is our best theoretical estimate which lies in the range 1.3-3.5  eV for different compounds. Interestingly vanadium based compounds show band gaps within the ideal visible range. The valence band maxima (VBM) and conduction band minima (CBM) mainly comprise of  N-p orbitals and M-d orbitals respectively.  The band gap increases as we go from V$\rightarrow$Nb$\rightarrow$Ta. Band structures show mostly flat bands at VBM and CBM, accounting for steep density of states at the band edges. A closer look at the band structure reveals mixture of dispersive and flat bands at the band edges, which may result in good thermoelectric performance. Band structure of the remaining compounds have similar features as these representative systems, and are shown in SM.\cite{supplement} \\
	
	{\bf  Evaluation for Renewable Energy Applications }
	\\ \\
	After analyzing the electronic structure data, we see that these compounds show favourable electronic structure to be considered as potential candidates for renewable energy applications. Below we evaluate these compounds for solar absorber, water-spitting and thermoelectric applications by carefully calculating few well-known descriptors.

\begin{figure*}[t]
		\centering
		\includegraphics[width=1\linewidth]{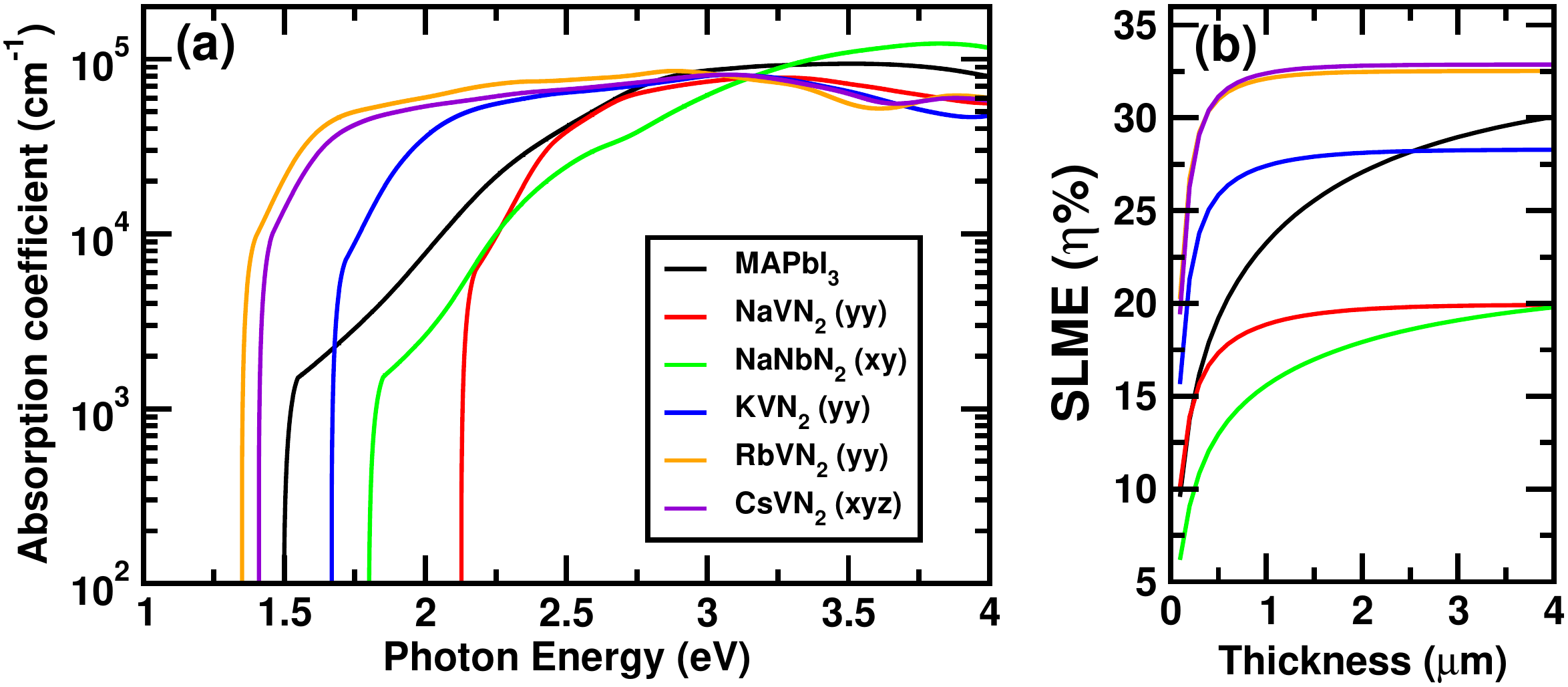}
		\caption{ For few promising nitrides, (a) Absorption coefficient and (b) SLME vs film thickness along the best possible crystal axes (along which the corresponding tensor component is maximum). Here yy means the contribution along y direction (yy component of related tensor), xy means the contribution in the xy plane (the average of xx and yy components, if xx, yy components are same) and xyz means isotropic (i.e. same contribution along all the three directions and hence average is same as any one of them). Similar data for polycrystalline sample (average of all three components irrespective of whether they are isotropic or not ) can be found in SM.\cite{supplement}}
		\label{fig:5}
	\end{figure*}

	{\bf Solar Absorber:}
	For a material to be considered as good solar absorber, it should posses few promising properties. $`$Spectroscopic limited maximum efficiency' (SLME) introduced by Yu. et al.\cite{yu2012identification} is one such parameter which quantifies those properties i.e. nature of band gap, optical absorption coefficient, etc, by providing a simulated efficiency parameter. It is a well-known descriptor, which predicts more realistic efficiency (e.g. for GaAs($\sim$28\%)\cite{yin2015superior}, MAPbI$_3$($\sim$30\%)\cite{yin2015superior}, CuInSe$_2$($\sim$28\%)\cite{yu2012identification}, CuGaSe$_2$($\sim$27\%)\cite{yu2012identification} etc..) than bare Shockley-Queisser parameter,\cite{shockley1961detailed}  (more details about SLME and related parameters can be found in Sec. S1-D(I) of SM).\cite{supplement} Figure~ \ref{fig:4}(b) shows SLME for all the twelve compounds at room temperature. The optical absorption is mainly dependent on two factors, one is optical transition strength, and the other is optical joint density of states (JDOS). The former can be calculated as the square of dipole matrix elements between the bands, and is plotted in Figure~ \ref{fig:3} alongside the band structures for three compounds. The transition strength is pretty high, which can be attributed to p- and d-orbital contributions at VBM and CBM respectively. Adding to that, here the VBM and CBM mainly show $`$d' and $`$p' orbital characteristics, which results in more flat bands near the band edges, hence likely to possess high JDOS.\cite{yin2015superior,heo2017cutas3}  All the properties favor excellent optical absorption, which is presented in Figure~ \ref{fig:5}(a) for five most promising AMN$_2$ compounds, whose SLME turn out to be $>$ 20 \%. The absorption coefficient of the state of the art photovoltaic material MAPbI$_3$ is also shown for comparison. Some of the proposed compounds, depending on their crystal structure show optical anisotropy, for which the respective maximum tensorial component (xx, yy, xy or xyz) of absorption coefficient and SLME are labelled in Figure~ \ref{fig:5}(a). These are the optical absorption contribution arising out of the dominant component from the dielectric function tensor. Here xx, yy or zz mean the contribution along x, y or z direction (xx, yy or zz component of related tensor) respectively, xy means in the xy plane (the average of xx and yy components, if xx, yy components are same) and xyz means isotropic (i.e., same contribution along the three directions). Except in case of layered compounds, the absorption coefficients in different directions are close for a given material. Similar simulated data for polycrystalline sample can be found in Fig.S6 in SM.\cite{supplement}  Although, all the compounds considered here have indirect band gaps, their optically allowed direct band gaps  are very close to the indirect ones (Table S3 in SM \cite{supplement}) and hence it does not affect the solar efficiency.

	RbVN$_2$(1.35 eV) and CsVN$_2$(1.41 eV) possess band gap in the ideal Shockley-Queisser region,\cite{shockley1961detailed} and show very sharp rise in the absorption spectra near band edge which results in very high short-circuit current density (J$_{sc}$) for these two. In order to get a practical idea, we have compared some device parameters and simulated efficiency with state of the art MAPbI$_3$. At lower thicknesses (about 500 nm) these two compounds posses J$_{sc}$ of 32.6 and 30.4 mA/cm$^2$ for RbVN$_2$ and  CsVN$_2$ respectively, which is  almost double than MAPbI$_3$ (16.8 mA/cm$^2$), resulting in almost 1.5 times higher SLME. Figure~ \ref{fig:5}(b) shows SLME vs. film thickness for five promising compounds along with a comparison with MAPbI$_3$. For RbVN$_2$ and  CsVN$_2$, at saturation thicknesses the current becomes 1.5 times, and simulated efficiency is at least 3-4\% higher. For the other three compounds, excellent absorption coefficient along with band gap in the good visible range shows good efficiency, especially for KVN$_2$ which shows SLME comparable to MAPbI$_3$. The other two compounds though do not possess SLME as high as the above three, but good photon absorption along with suitable band gap values make them potential candidates for solar water splitting applications, which we discuss next.

	{\bf  Water splitting: }
	Photochemical (PC) and photoelectrochemical (PEC) water splitting are the scientific processes to split water to produce hydrogen gas using solar energy.\cite{yao2018photoelectrocatalytic,  jafari2016photocatalytic} Using hydrogen as fuel is one of the cleaner, sustainable, renewable ways to meet the global energy need. In this process, mainly two reaction occurs (i) hydrogen evolution reaction (HER):  $2H^{+}+2e^{-} \rightarrow H_2$   and (ii) oxygen evolution reaction (OER):  $2H_2O \rightarrow  4H^+ + O_2 + 4e^-$. For a material to be considered as photocatalyst for both the reactions, its VBM needs to be below (more positive than) the oxidation potential of O$_2$/H$_2$O, i.e 1.23 V vs reversible hydrogen electrode (RHE) and CBM needs to be above (more negative than) the reduction potential of H+/H$_2$, i.e. i.e.,0.00 V vs RHE. Ever since the groundbreaking discovery of Fujishima and Honda, where they reported first photocatalytic water splitting using TiO2,\cite{fujishima1972electrochemical} mostly oxides are considered as catalysts due to their stability and easy synthesis.\cite{castelli2012computational} But oxides having band gap in the UV region could not utilise the visible solar irradiation which consists most of the solar power. Recently, various high-throughput studies have been reported considering metal oxides, oxynitrides, etc as possible electrode materials for water splitting.\cite{castelli2012computational, castelli2012new, sawada2018high, wu2013first} But in the practical scenario, we are still much behind the target efficiency to make this commercially viable. This calls for the need to discover novel materials for this purpose and study their potentiality. Recently, metal nitrides have emerged as candidate materials for photo(electro)chemical water splitting, showing better solar absorption, charge transport properties, etc.\cite{zakutayev2016design, han2018recent} Nitrogen being less electronegative than oxygen, N-2p orbitals sit above O-2p orbitals, pushing the VBM (which mainly consists of anion-2p in both cases) higher and thus making the overall band gap lower for nitrides than oxides, resulting in better utilisation of solar spectrum. Also nitrides are still less explored, which provides a great opportunity for new materials discovery.

	\begin{figure*}[t]
		\centering
		\includegraphics[width=0.95\linewidth]{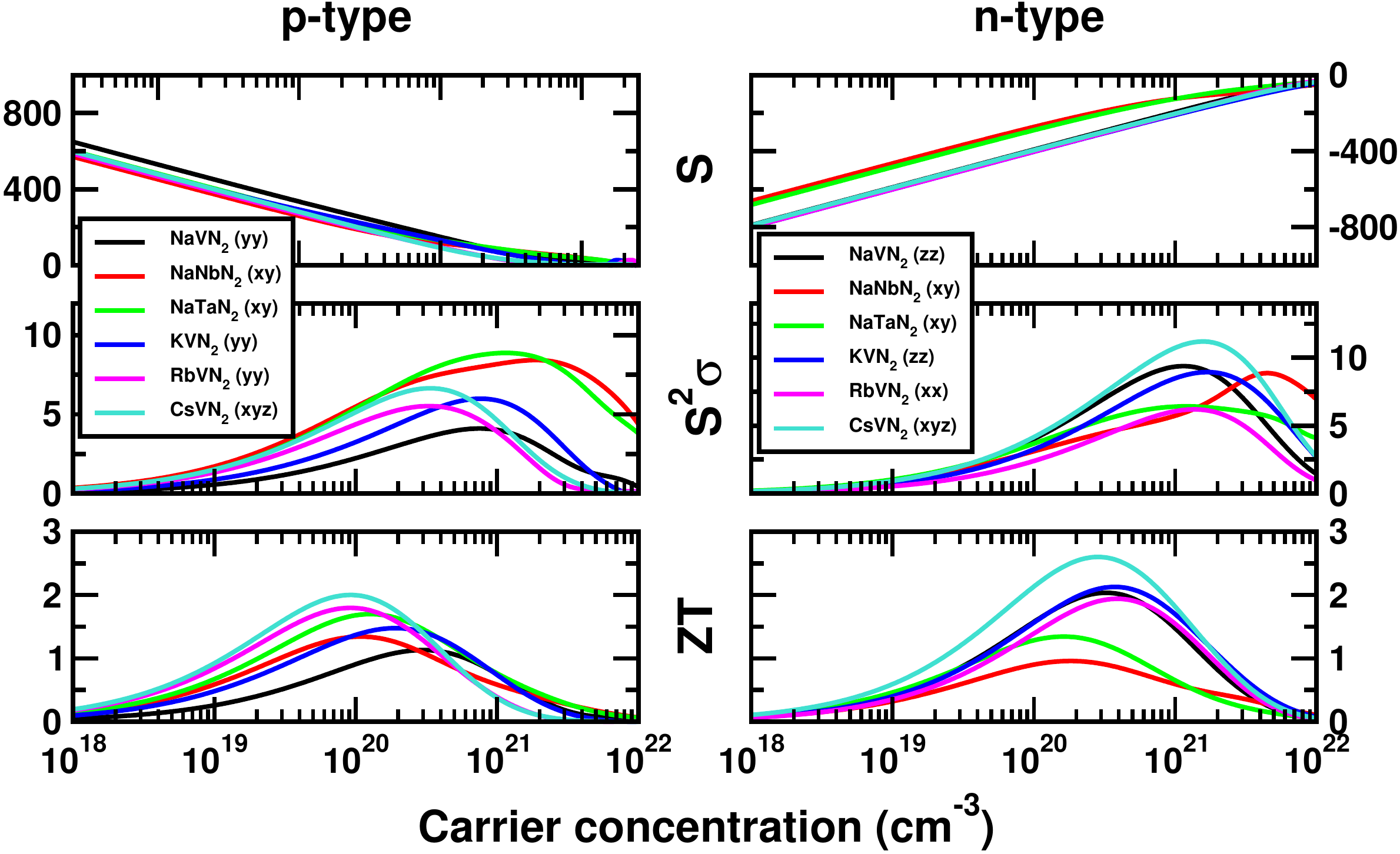}
		\caption{Seebeck coefficient (S) (in $\mu VK^{-1}$), Power factor ($S^2\sigma$) (in $mWm^{-1}K^{-2}$) and Figure of merit (ZT) for some of the promising nitrides along best possible crystal axis orientation. Here xx, yy, zz mean the contribution along x, y, z direction (xx, yy, zz component of related tensor) respectively, xy means in the xy plane (the average of xx and yy components, if xx, yy components are same) and xyz means isotropic.}
		\label{fig:6}
	\end{figure*}
	
	From the above discussion, it is clear that for a semiconducting material to be considered for light harvesting in PEC water splitting cell, it needs to have some suitable properties, such as (i) robust stability, (ii) suitable band gap ($E_g$) (iii) high optical absorption, (iv) band edges well positioned relative to water redox levels, and v) good carrier mobility. As for our proposed ternary nitrides, we have already discussed about the stability in the previous section. Our calculation using G$_0$W$_0$ method shows some of the material to possess ideal band gaps to be considered for water-splitting. They also show strong optical absorption properties suitable to achieve high solar to hydrogen efficiency. Simulated band edge effective mass values indicate possibility of good carrier mobility.(Table S3 in SM \cite{supplement})  The next important thing to determine is the band edge positions for these materials with respect to OER and HER. We  have used the empirical model proposed by Butler and Ginley,\cite{butler1978prediction}  to determine these band edge positions. The details about this model is given in SM.\cite{supplement}
	
	Figure~ \ref{fig:4}(c) shows the band edge positions of all the twelve nitrides. It is clear that all the nitrides have well positioned CB edge, to be used as photocathode for HER in a PEC cell. However, by looking at the band gap and absorption properties, we propose ternary vanadium nitrides along with the already synthesized NaNbN$_2$ \cite{jacobs1993synthesis2} and NaTaN$_2$\cite{jacobs1989synthese} to be most suitable photocathodes. On the other hand, Tantalum nitrides (most of which are already experimentally synthesized\cite{jacobs1989synthese}) along with RbNbN$_2$ can be suitable candidates to work as a medium where HER and OER can occur simultaneously.\\

	{\it Overcoming photocorrosion issues in nitride photocatalysts : }	 The main problem which may arise  with nitrides is the inherent photocorrosion i.e. instability in aqueous solution under illumination. The occurrence of such corrosion lies in the instrinsic anion orbital characteristics.\cite{su2017stability} Over the years, researchers are working to overcome these difficulties, as such some ways are proposed. One is, physical protection of unstable surface, e.g Li et al. used ferrihydrite to protect unstable Ta$_3$N$_5$ \cite{liu2014tantalum} and related research.\cite{hu2015thin, scheuermann2016atomic} The other methods are using co-catalysts, surface engineering, etc. which are discussed in details elsewhere.\cite{yang2015enabling, kibria2016atomic} These techniques which mainly depend on the physical and chemical properties of the associated compounds can help overcome the difficulties using nitrides and make them reach the desired potentials.

	{\bf Thermoelectrics: }
	Thermoelectrics (TE) offer an environmental friendly way to convert heat into electricity, thus a valuable contributor to the waste heat recovery and renewable energy generation.\cite{he2017advances} From electronic structure point of view,  most thermoelectric materials  possess small band gap ($<$\ 1eV). Wider gap ($>$\ 1eV) materials, however, can also show promising thermoelectric efficiency, given they show some favourable structural and electronic properties.\cite{gorai2017computationally, zeier2016thinking, morelli2008intrinsically}
	
	To quantify the performance of a thermoelectric material, there is a well known  thermoelectric figure of merit, $$ ZT=\frac{S^2\sigma }{\kappa}T$$ where S is Seebeck coefficient, T is corresponding temperature, $\sigma$ and $\kappa$ are the electrical and thermal conductivity of the material respectively. $\kappa$ consists of two parts, electronic ($\kappa_e$), and lattice ($\kappa_L$) contribution. According to Weidemann-Franz law,  $\kappa_e$ is related to electrical conductivity $\sigma$ by, $\kappa_e=L\sigma T$ where $L$ is Lorenz number. L can be estimated within a reasonable accuracy via the  equation $$L=1.5+ exp|\frac{-|S|}{116}|$$ where $S$ is in units of $\mu V/K$.\cite{kim2015characterization}  We calculate $\kappa_L$ using Cahill's model\cite{cahill1992lower} (details in supplement\cite{supplement}) which estimates lattice contribution to the amorphous limit for a  given material. The rest of the quantities, S, $\sigma$, are calculated via solving the semiclassical Boltzmann transport equation, as implemented in BoltzTraP code, \cite{madsen2006boltztrap} within the constant relaxation time approximation.\cite{zhu2015computational} We discuss more about the interdependence of these parameters in SM.\cite{supplement}	
	
	All of the proposed nitrides possess favourable band structures to be considered for thermoelectric applications. We first estimate maximum possible ZT for all the compounds at $600$ K as shown in Figure~ \ref{fig:4}(d). The lowest band gap for these nitrides is 1.35 eV, which ensures negligible bipolar thermal conductivity. Doping is a necessary step to achieve good TE performance and higher the band gap, harder it gets to intentionally dope a material. Depending on the performance (max. ZT) and band gap values, we have identified six potential TE materials, the detailed properties of which (S, power factor (S$^2\sigma$) and ZT) are shown in Figure~ \ref{fig:6}. The description of various tensorial components (xx, yy, zz, xy and xyz) are same as those in Fig. \ref{fig:5}. )A brief discussion on relevant dopants can be found in sec. S1.D.III in SM \cite{supplement}.)

	A close analysis of the band structure of CsVN$_2$ (Figure.\ref{fig:3}(a)) reveals the presence of both flat and dispersive nature of bands along different {\bf{k}} directions. This kind of band structures are known as pudding mold type bands, and known to show good thermoelectric properties.\cite{isaacs2019remarkable} For CsVN$_2$ we see this kind of bands in both VBM and CBM, resulting in a very high Seebeck coefficient, as obvious from Figure~ \ref{fig:6}. Also the VBM resides at a point other than $\Gamma$, which indicates higher valley degeneracy thus better performance. Out of the other five, NaNbN$_2$ and NaTaN$_2$ are already reported\cite{ohkubo2015anisotropic} as good thermoelectrics, which is confirmed by our calculations. These materials show secondary bands pretty close to VBM and CBM (as can be seen in Fig. S2 of SM\cite{supplement}) which can effectively increase the valley degeneracy with increase in temperature, thus giving excellent thermoelectric properties. The other three compounds (NaVN$_2$, KVN$_2$ and RbVN$_2$) also possess similar kind of band structures and thus turn out to be equally promising, with achievable ZT value in the range of 2 or higher [see SM \cite{supplement} for details]. Lastly, our calculated Electronic Fitness Function (EFF)\cite{xing2017electronic} (for details, see section S1(D.III) and  S2(H) of SM.\cite{supplement}) for these compounds are comparable to well known TE compounds like PbTe, ZrNiSn, Mg$_3$Sb$_2$, GeTe,  etc. This suggests that the typical band shape favours high electronic performance,  as also confirmed by high values of power factor (S$^2\sigma$).\cite{putatunda2018thermoelectric} 
	
	\begin{figure}[t]
		\centering
		\includegraphics[width=1\linewidth]{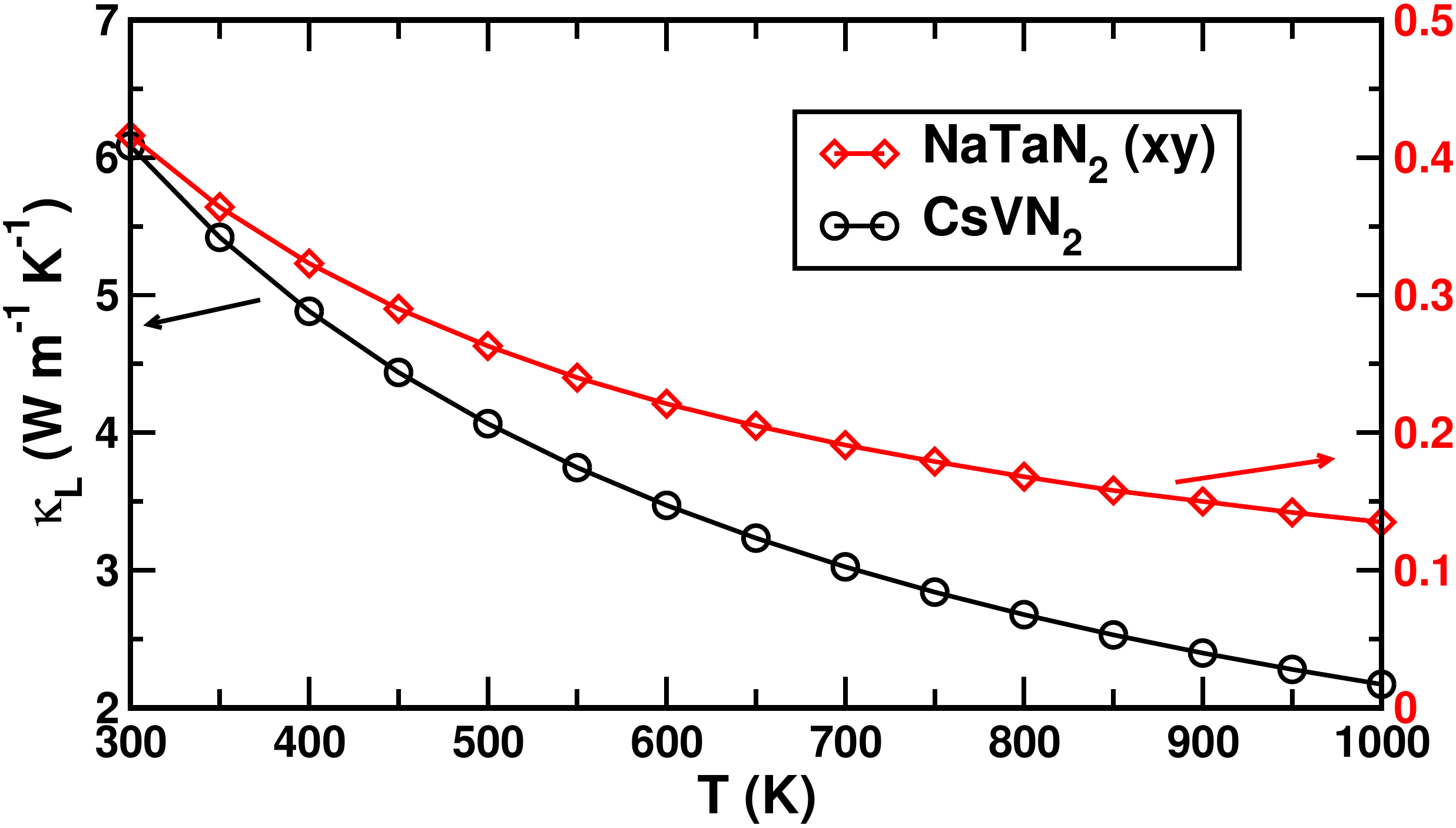}
		\caption{Lattice thermal conductivity vs temperature for NaTaN$_2$ and CsVN$_2$.}
		\label{fig:7}
	\end{figure}
	
	One of the key factors, which often plays a crucial role in optimizing TE performance of a material is lattice part of the thermal conductivity ($\kappa_L$). To dig in more details, we have simulated the lattice thermal conductivity for two of the compounds, namely NaTaN$_2$ and CsVN$_2$, by solving the Boltzmann transport equation for phonons as implemented in phono3py code.\cite{togo2015distributions, mizokami2018lattice} This method includes anharmonic component by calculating the third order force constants, and is known to be optimal for accurately calculating $\kappa_L$.\cite{plata2017efficient}
	Figure~ \ref{fig:7} shows the temperature dependence of $\kappa_L$. We see that NaTaN$_2$ shows ultralow $\kappa_L$ which is seen for some layered compounds\cite{kumar2016lattice} and makes it highly suitable for thermoelectric applications, even at lower temperatures. CsVN$_2$ also shows moderate to low $\kappa_L$, making it suitable for moderate to higher temperature applications.   We expect similar order of magnitude of $\kappa_L$ for other four compounds as well.

	{\bf  Conclusion:}
	In summary, we performed a careful investigation of previously unexplored ternary alkali metal pnictides AMX$_2$ from the renewable energy application perspective. We have done a thorough stability analysis via considering most probable decomposition pathways, and found twelve stable nitride compounds. Next,  a detailed electronic structure analysis is carried out by using most accurate exchange correlation functionals. Most of these compounds show favourable band structures and other electronic properties to be considered for various renewable energy applications such as photovoltaic absorber, photo(electro) chemical water splitting and thermoelectrics. CsVN$_2$ and RbVN$_2$, having band gap in the ideal S-Q region with sharp rise in absorption spectra near band edge turn out to be potential candidates for solar absorbers with simulated efficiency much higher than the state of the art MAPbI$_3$. Most of the compounds show favourable band edge positions (with respect to water redox level) either to be used as photocathodes, or both photoelectrodes for solar water splitting devices. Careful analysis of the band structure for CsVN reveals a mixture of dispersive and flat band along different {\bf{k}}-directions, showing good thermoelectric performance. Our simulated lattice thermal conductivity confirms high achievable ZT for CsVN$_2$ and NaTaN$_2$, even at moderate to low temperatures. Some of the other compounds also show favorable band structures which results in high figure of merit.  Experimental synthesis of some of these compounds are previously reported, but detailed properties remain unexplored. We strongly believe that computational efforts, such as the present report, can direct the experimentalists towards realistic targets and speed up the discovery of novel materials. It is highly desirable to experimentally synthesize the proposed compounds here and verify their functional properties.

	{\bf  Methods:}
	First principles calculations were performed using Density Functional Theory (DFT)\cite{kohn1965self} as implemented in Vienna Ab-initio Simulation Package (VASP).\cite{kresse1996efficiency,kresse1999ultrasoft} Thermoelectric transport  parameters are calculated using BoltzTraP code.\cite{madsen2006boltztrap} For lattice thermal conductivity calculation, we have used phono3py code.\cite{togo2015distributions, mizokami2018lattice}  More details about the procedure and validation can be found in Sec. S1 of SM.\cite{supplement}\cite{ blochl1994projector, dudarev1998electron, lan2018linear, xu2015accurate, perdew2008restoring, zhao2017design, faghaninia2015ab, mouhat2014necessary, hill1952elastic,  green2004third, xu2000absolute, castelli2015new, trasatti1990surface, castelli2013stability, pei2011convergence, sun2017thermoelectric,  yan2015material, tang2015convergence}

	{\bf Acknowledgements:}
	J.K. acknowledges financial support from IIT Bombay in the form of a Teaching Assistantship. AA acknowledges National Center for Photovoltaic Research and Education (NCPRE), IIT Bombay for possible funding to support this research.

\end{document}